\documentclass[preprint,aps]{revtex4}
\usepackage{graphicx}
\usepackage{color}
\usepackage{amsmath}
\usepackage{float}

\begin{document}

\title{Self modulated dynamics of a relativistic charged particle beam in plasma wake field excitation}

\author{T. Akhter$^{a}$, R. Fedele$^{a}$, S. De Nicola$^{b,a}$, F. Tanjia$^{a}$, D. Jovanovi\'c$^{c}$, and A. Mannan$^{d}$,}
\address{$^a$ Dipartimento di Fisica, Universit\`a di Napoli Federico II and INFN Sezione di Napoli, Italy}
\address{$^b$ CNR-SPIN, Complesso Universitario di Monte S'Angelo, Napoli, Italy}
\address{$^c$ Institute of Physics, University of Belgrade, Belgrade, Serbia}
\address{$^d$ INFN Sezione di Napoli, Napoli, Italy}

\renewcommand{\abstractname}{}
\begin{abstract}
Self modulated dynamics of a relativistic charged particle beam is
reviewed within the context of the theory of plasma wake field excitation. The self-consistent
description of the beam dynamics is provided by coupling the Vlasov
equation with a Poisson-type equation relating the plasma wake potential to the beam density.
An analysis of the beam envelope self-modulation is then carried out and the
criteria for the occurrence of the instability are discussed thereby.
\end{abstract}

\maketitle
\section{INTRODUCTION}
\noindent It is well known that ultra short intense accelerating gradient, to conceive novel scheme of particle acceleration, can be achieved by launching very high intense ultra short laser pulses or relatively intense relativistic charged particle beams, in a plasma. Both the laser pulses and the charged particle bunches, with different mechanisms, drive a large amplitude plasma wave that propagates behind them as a water wake propagates behind the motor boat. The mechanism which employs laser pulses, is usually referred to as the laser wake field (LWF) excitation, since the plasma wake is produced by the ponderomotive effect and propagates behind the laser pulse with phase velocity equal to laser group velocity \cite{Rosenbluth-Liu,Tajima1979,Gorbunov1987,Sprangle1988}. When the charged particle bunches are employed, another mechanism, called plasma wake field (PWF) excitation, takes place due to the local charge neutrality violation introduced by the bunches, and due to the current that the bunches themselves introduce in the plasma. Here the plasma wake propagates at the same speed of the bunch. If the bunch, i.e., beam, is sufficiently long compared to a plasma wavelength (i.e., the bunch length $\sigma_z$ is much greater than the plasma wavelength $\lambda_p$), the beam experiences the self-consistent PWF interaction \cite{Chen1985,Ruth1985,Chen1986,Rosenzweig1987}. The main effect caused by this interaction is the self-modulation of the beam \cite{5-1,5-2} that may lead the beam to the unstable conditions, such as the self-modulation instability (SMI) \cite{Schroeder2012a,Muggli2012}. Recently, PWF mechanism has been used to describe the self-modulation of longitudinal and transverse beam dynamics in both overdense (the beam density is much smaller than the plasma density) and underdense (the beam density is nearly equal or a little above the plasma density) regimes \cite{Fedele2014n,Fedele2014c,Fedele2014a,Fedele2014b}.

In this manuscript, we are going to review some recent investigations on self-modulation instabilities within the context of PWF excitation in overdense regime. For sufficiently long beams, the longitudinal beam dynamics can be neglected, therefore we focus on the transverse dynamics only \cite{Fedele2014n,Fedele2014c,Fedele2014a,Fedele2014b,Fedele2012}. However, the motion in the transverse directions (i.e., $x$ and $y$), is non relativistic but each particle of the beam has the mass that is increased by the gamma factor, i.e., $m_0\gamma_0$ where $m_0$ is the rest mass and $\gamma_0$ is the unperturbed relativistic single particle \textit{gamma factor}.

\noindent We assume that a relativistic charged particle beam enters the plasma of unperturbed density $n_0$. The beam travels along the $z$ direction with velocity $\beta c$ ($\beta \approx 1$ and $c$ being the speed of light), energy $E_0 = m_0\gamma_0 c^2$. We also assume that the plasma is cold and strongly magnetized by an uniform and constant magnetic field of magnitude $B_0$ oriented along $z$. Then, the linearized set of Lorentz-Maxwell's equations associated with the ``beam + plasma" system can be reduced to the following Poisson-like equation \cite{Chen1986,Fedele2014a,Fedele2014b}:
\begin{equation}\label{rn1}
\left(\nabla_\bot^2 - k_s^2\right)U_w = k_s^2\frac{\rho_b}{n_0\gamma_0},
\end{equation}
where $\mathbf{r}_\perp$ is the transverse component of the position vector and $\xi = z-\beta c t$ ($t$ being the time), $\rho_b = \rho_b(\mathbf{r}_\bot,\xi)$ is the local beam density, $U_w\left(\mathbf{r}_\bot,\xi\right)=q(A_{1z}-\phi_1)
/m_0\gamma_0c^2$ is the wake potential energy normalized with respect to $m_0\gamma_0 c^2$, $\phi_1$ and $A_{1z}$ are the perturbations of the scalar potential and the longitudinal (i.e., parallel to $z$) component of the vector potential, respectively. Here,
$k_s=k_{p}^2/k_{uh}$,\,$k_{uh}=\omega_{uh}/c$,\, $k_{c}=\omega_{c}/c$\,,
$\omega_c$ and $\omega_{uh}=(\omega_p^2+\omega_c^2)^2$ are the electron cyclotron frequency and the upper hybrid frequency, respectively, and $k_p = \omega_p/c$ ($\omega_p$ is the plasma frequency). Note that Eq. (\ref{rn1}) has been obtained for the electrostatic condition in the co-moving frame. Then, the self consistent description has been obtained by providing Vlasov equation for the beam dynamics (see next Section).

\section{SELF-CONSISTENT SYSTEM IN CYLINDRICAL SYMMETRY}
We assume the cylindrical symmetry to describe the self-modulated dynamics of beam-plasma system. Therefore, the spatio-temporal phase-space evolution of a collision-less beam in the presence of the strong magnetic field $B_0$ is provided by the following Vlasov equation in terms of cylindrical coordinate (for details se Refs. \cite{Fedele2014a,Fedele2014b}):
\begin{equation}\label{rn2}
  \frac{\partial f}{\partial \xi} + p\frac{\partial f}{\partial r} - \frac{\partial V}{\partial r}\frac{\partial f}{\partial p}=0\,,
\end{equation}
where
\begin{equation}\label{rn2a}
  V(r,\xi) = \frac{1}{2}Kr^2 + U_w(r,\xi),
\end{equation}
$f(r, p, \xi)$ is the normalized phase space distribution function, $r$ is the cylindrical radial coordinate and $p$ is the corresponding conjugate momentum. Here
\begin{equation}\label{rn3}
  K=\left(\frac{k_c}{2}\right)^2=\left(-\frac{qB_0}{2m_0\gamma_0c^2}\right)^2\,.
\end{equation}
In the case of the strongly nonlocal regime (i.e., the beam spot size is much smaller than the plasma wavelength), we can simplify the Poisson-like equation that relates the beam density to the wake potential. Then, in cylindrical symmetry, Eq. (\ref{rn1}) in strongly nonlocal regime (i.e., $\left|\nabla_\perp^2 U_w\right|\gg k_s^2 \left|U_w\right|$) becomes
\begin{equation}\label{rn4}
  \frac{1}{r}\frac{\partial}{\partial r} \left(r \frac{\partial U_w}{\partial r}\right)= k_s^2 \,\lambda_0\,\, \rho_b,
\end{equation}
where $\lambda_0 = N/n_0 \gamma_b \sigma_z$ ($\sigma_z$ being the beam length), and the beam density is given as,
\begin{equation}\label{rn5}
\rho_b(r,\xi) = \frac{N}{\sigma_z}\,\int f(r,p,\xi)\,d^2p\,,
\end{equation}
where $N$ is the total number of beam particles.
\section{ANALYSIS OF SELF-MODULATED MOTION}
In the nonlocal regime, the potential $ V(r,\xi)$ has a minimum for $r = 0$ and the beam is confined around the z-axis. Therefore, we can expand $V$ in aberration-less approximation, as,
\begin{equation}\label{rn6a}
  V(r,\xi) \simeq V(0,\xi) + \left(\frac{\partial V}{\partial r}\right)_{r=0}r+\frac{1}{2}\left(\frac{\partial^2 V}{\partial r^2}\right)_{r=0}r^2.
\end{equation}
Equations (\ref{rn2a}) and (\ref{rn4}) indicates that $V(0,\xi)= U_w(0,\xi)=0$ and $\left(\partial V/\partial r\right)_{r=0}=0$. Hence, the potential reduces to,
\begin{equation}\label{rn6b}
  V(r,\xi)\simeq \frac{1}{2}\,k(\xi)\,r^2\,,
\end{equation}
where
\begin{equation}\label{rn6c}
  k(\xi)=\left(\frac{\partial^2 V}{\partial r^2}\right)_{r=0} \equiv  K+\overline{K}(\xi) \,,
\end{equation}
and
\begin{equation}\label{rn6d}
  \overline{K}(\xi)= \pi k_s^2\lambda_0\,\int_{0}^\infty f(0,p,\xi)\,p\,dp\,.
\end{equation}
The condition for having minimum at $r=0$ implies that $k(\xi) > 0$ which consequently indicates the condition, $K > - \overline{K}(\xi)$.
For an initial Gaussian beam distribution and for potential given by Eq. (\ref{rn6b}), the solution of the Vlasov equation (\ref{rn2}) can be written as,
{\small\begin{equation}\label{rn6}
  f(r,p,\xi)=A\,\exp\left\{-\frac{1}{\epsilon}\left[\gamma(\xi)\,r^2+
  2\,\alpha(\xi)\,rp+\beta(\xi)\,p^2 \right]\right\},
\end{equation}}
where $A$ is a normalization constant, $\epsilon$ is the beam emittance and the functions $\alpha(\xi)$, $\beta(\xi)$ and $\gamma(\xi)$ (the analog of the Twiss parameters in electron/radiation beam optics) obey the following first-order ordinary differential equations, viz:
\begin{equation}\label{rn7}
  \frac{d\gamma}{d\xi}=2k\alpha,\,\,\frac{d\beta}{d\xi}=-2\alpha\,\,\text{and}
  \,\,\,\,\frac{d\alpha}{d\xi}=-2\gamma + 2k\beta\,.
\end{equation}
Consequently, Eqs. (\ref{rn6}) and (\ref{rn7}) allow us to finally get
the following envelope equation, viz.,
\begin{equation}\label{rn8}
  \frac{d^2 \sigma_\bot}{d\xi^2}+ K\,\sigma_\bot + \frac{\eta}{\sigma_\bot} -\frac{\epsilon^2}{\sigma_\bot^3}=0\,,
\end{equation}
where $\epsilon^2= \sigma_\bot^2\,\sigma_{p\bot}^2-\sigma_{rp}^2$, $\sigma_{\bot}=\,\langle r_\bot^2 \rangle^{1/2}$ is the transverse beam spot size, $\sigma_{p_\bot} = \langle p_\bot^2 \rangle^{1/2}$ is the momentum spread, and $\sigma_{rp}=\,\langle r p \rangle$ is the correlation moment. Here, $\eta=k_s^2\lambda_0/2\pi=k_s^2 N/2\pi\sigma_z n_0\gamma_0$, which accounts for the collective effects.\\

In order to perform an analysis of the self-modulation due to the PWF interaction in the strongly nonlocal regime and aberration-less approximation, we use the method of Sagdeev potential that can be constructed from the envelope equation (\ref{rn8}), given as,
\begin{equation}\label{rn9}
\frac{d^2\widetilde{\sigma}}{d\widetilde{\xi}^2}=-\frac{\partial V_s(\widetilde{\sigma})}{\partial\widetilde{\sigma}}\,,
\end{equation}
where
\begin{equation}\label{rn10}
V_s(\widetilde{\sigma})=\frac{1}{2}\widetilde{K}\widetilde{\sigma}^2+\frac{1}{2}\widetilde{\eta}\ln\widetilde{\sigma}^2+
\frac{1}{2\widetilde{\sigma}^2}\,.
\end{equation}
Here we have introduced the dimensionless quantities:
$\widetilde{\xi}=\xi\epsilon/\bar{\sigma}^2$\,,\,\,\,$\widetilde{\sigma}= \sigma_\bot/\bar{\sigma}$\,,\,\,\,$\widetilde{K}=K\bar{\sigma}^4/\epsilon^2$\,,\,\,\,$\widetilde{\eta}=
\eta\bar{\sigma}^2/\epsilon^2$\,, $\bar{\sigma}$ is the real positive root of the equation,
\begin{equation}\label{rn11}
 K\,\bar{\sigma}+ \frac{\eta}{\bar{\sigma}} -\frac{\epsilon^2}{\bar{\sigma}^3}=0\,.
\end{equation}
We can use Eqs. (\ref{rn9}) and (\ref{rn10}) to analyze self-modulated motion of the system including stability and instability.
\begin{figure}[h!]
  \includegraphics[width=80mm]{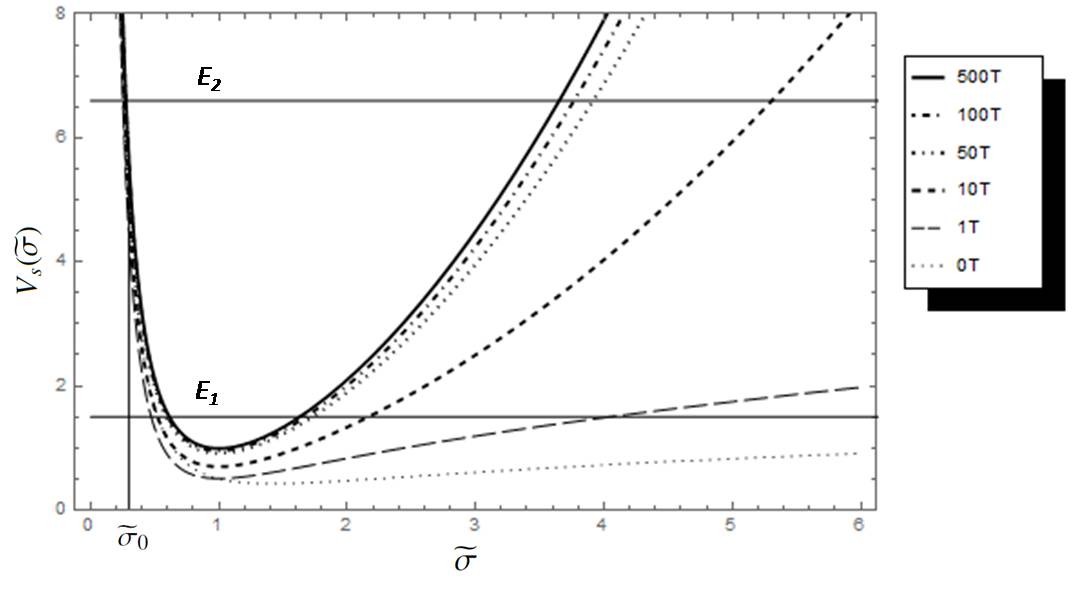}
  \caption{Sagdeev potential $V_s$ as defined by Eq. (\ref{rn10}) for the different values of the external magnetic field amplitude $B_0$ and $\widetilde{\eta}$ (calculated for each $B_0$ according to Eq. (\ref{rn11})). Here, $\epsilon = 10^{-7} rad\,m$, $\widetilde{\sigma}_0=1.5$, $\left(\frac{d\widetilde{\sigma}}{d\widetilde{\xi}}\right)_{\widetilde{\xi}=0} =0$.}\label{F1}
\end{figure}
\begin{figure}[h!]
  \includegraphics[width=80mm]{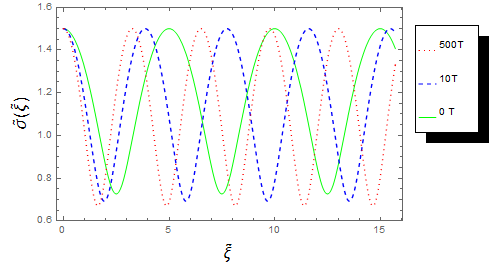}
  \caption{Numerical solutions of the Eq. (\ref{rn8}) for different values of external magnetic field  and $\widetilde{\eta}$. Here, $\epsilon = 10^{-7} rad\,m$, $\widetilde{\sigma}_0=1.5$, $\left(\frac{d\widetilde{\sigma}}{d\widetilde{\xi}}\right)_{\widetilde{\xi}=0} =0$.}\label{F2}
  \end{figure}
For the initial conditions $\widetilde{\sigma}_0\equiv\widetilde{\sigma}(\widetilde{\xi}=0) = \sigma_0/\bar{\sigma}$ and $\left(d\widetilde{\sigma}/d\widetilde{\xi}\right)_{\widetilde{\xi}=0}\,$ the mechanical energy of a
representative point in the \textit{$\widetilde{\sigma}$ - space} is fixed, i.e.,
\begin{eqnarray}
&&\hspace{-1.3cm}E = \frac{1}{2}\left(\frac{d\widetilde{\sigma}}{d\widetilde{\xi}}\right)^2+
\frac{1}{2}\widetilde{K}\widetilde{\sigma}^2+\frac{1}{2}\widetilde{\eta}\ln\widetilde{\sigma}^2+
\frac{1}{2\widetilde{\sigma}^2} \nonumber\\
&&\hspace{-1.1cm}=\frac{1}{2}\widetilde{\sigma'}_0^2+
\frac{1}{2}\widetilde{K}\widetilde{\sigma}_0^2+\frac{1}{2}\widetilde{\eta}\ln\widetilde{\sigma}_0^2+
\frac{1}{2\widetilde{\sigma}_0^2}\,=\,\mbox{constant}\,.
\end{eqnarray}
Figures \ref{F1} and \ref{F2} show the variations of $V_s(\widetilde{\sigma})$ and corresponding numerical solutions of the envelope equation for given initial conditions ($\widetilde{\sigma}_0=1.5$, $\left(d\widetilde{\sigma}/d\widetilde{\xi}\right)_{\widetilde{\xi}=0}=0\,$) and different parameters $\widetilde{K}$ and $\widetilde{\eta}$.
\section{QUALITATIVE STABILITY ANALYSIS}
The analysis of the displayed plots leads to conclude that, during the beam evolution, provided that ideally the system is always in the strongly nonlocal regime, the self-modulation is stable and the potential well is always trapping. However, as we have already indicated above, this condition is fulfilled provided that $\left|\nabla_\perp^2 U_w\right|\gg k_s^2 \left|U_w\right|$ (i.e., $k_s \sigma_\bot \ll 1$). Therefore, for a suitable choice of the initial conditions, such that $\sigma_\bot$ satisfies this inequality, Figures \ref{F1} and \ref{F2} show that the envelope self modulations are stable and represents periodic solutions in $\sigma_\bot$. This corresponds to fix the total energy to relatively small values, such as $E_1$ in Figure \ref{F1} for some of the profiles of the Sagdeev potential. If the energy $E$ coincides with the minimum of the potential well (i.e., $\widetilde{\sigma}_0 = 1$), the beam is in a stationary state (self-equilibrium) and its spot size does not change, being fixed to the initial value $\widetilde{\sigma}_0$ (note that $\left(d\widetilde{\sigma}/d\widetilde{\xi}\right)_0=0$), according to Eq. (\ref{rn11}). If the energy is fixed to a value above the minimum of the potential well with $\widetilde{\sigma}_0 >1$ and, for simplicity, $\left(d\widetilde{\sigma}/d\widetilde{\xi}\right)_0=0$, the motion of the representative point will be a nonlinear oscillation, provided that the condition $k_s\sigma_\bot\ll 1$ is still fulfilled. In particular, values of $E$ that are very close to the minimum imply harmonic oscillations of $\sigma_\bot$ (see Figure \ref{F2}). However, as $B_0$ increases, sensitive changes in both the structure of the Sagdeev potential and the period of the corresponding beam self-modulation are observed in such a way that the beam spot size will execute more and more stable nonlinear transverse oscillations.
\begin{figure}[h!]
  \includegraphics[width=75mm]{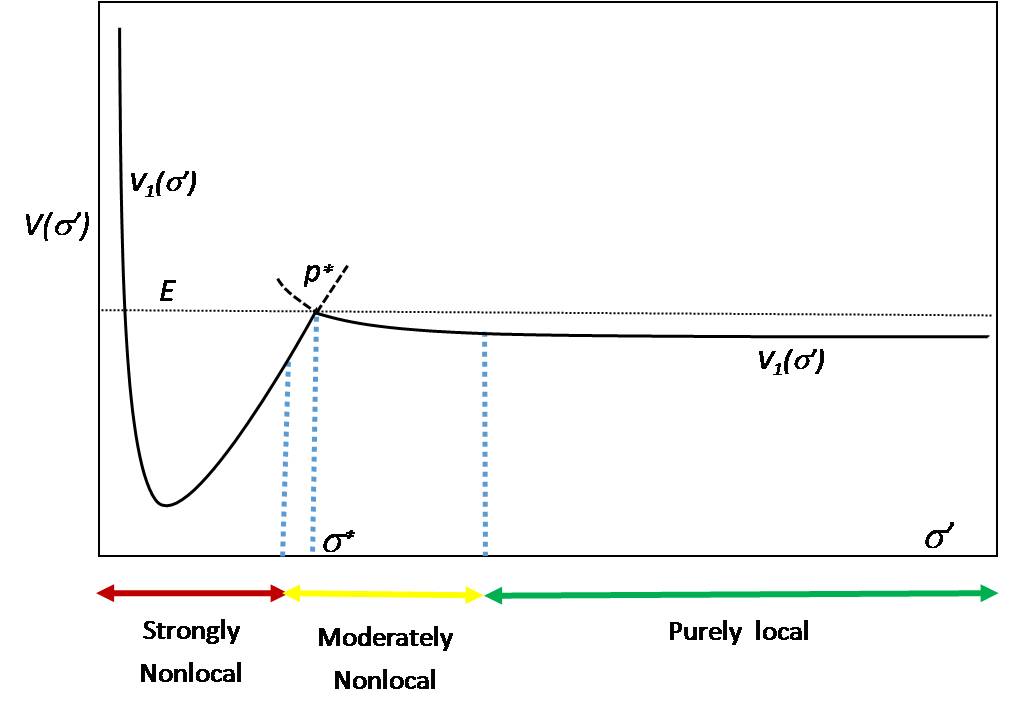}
  \caption{Qualitative plot of $V$ over all the three different regimes.}\label{F3}
\end{figure}
It is worthy noting that, although the condition $k_s\sigma_\bot\ll 1$ is initially fulfilled, there are specific initial conditions that lead to an unstable evolution of the beam modulation, i.e., the self-modulation instability (SMI). In fact, as illustrated in Figure \ref{F1}, if we choose $\sigma_0$ sufficiently less than $\bar{\sigma}$ (i.e., $\widetilde{\sigma}_0$ sufficiently less than $1$), the corresponding energy $E$ will be sufficiently high to allow $\widetilde{\sigma}$ to grow until the kinetic energy of the representative point becomes zero, i.e., $1/2\,(d\widetilde{\sigma}/d\widetilde{\xi})^2=0$.  Then, if the magnetic field is strong enough, its trapping effect limits the excursion of $\widetilde{\sigma}$ oscillations to values that are still of the same order of magnitude of $\widetilde{\sigma}_0$ (Figure \ref{F1}).
\begin{figure}[h!]
  \includegraphics[width=75mm]{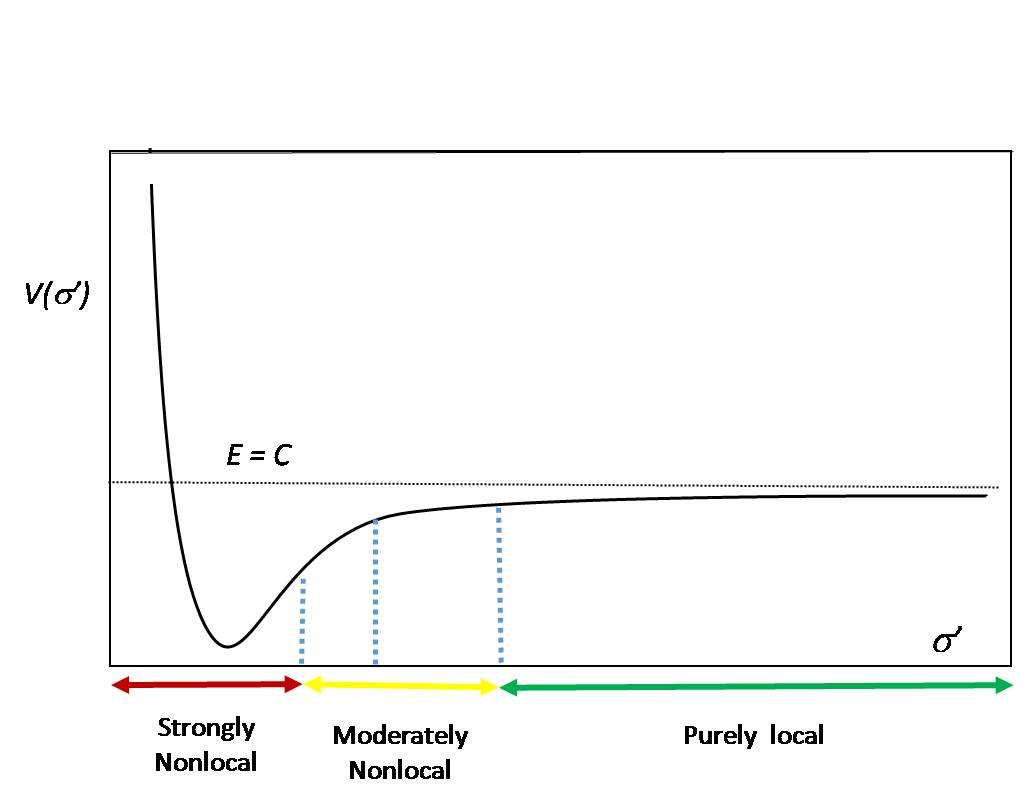}
  \caption{Qualitative plot of $V$ over all the regimes with a qualitative smoothing of the moderately non-local regime.}\label{F4}
\end{figure}
However, if $B_0 =0$ or assumes relatively small values, $\widetilde{\sigma}$ can easily grow and becomes muche greater than $\widetilde{\sigma}_0$. But, when this  is just about to happen, the condition $k_s \sigma_\bot <<1$ is no longer fulfilled and, therefore, the system enters progressively into the \textit{moderately nonlocal regime} (i.e., $k_s\sigma_\bot \sim 1$) and then into the \textit{purely local regime} (i.e., $k_s\sigma_\bot >>1$) \cite{Fedele2014a}. We conclude that, although the system starts with an initial condition falling into the strongly nonlocal regime, it may rapidly violate the condition $k_s \sigma_\bot <<1$ in such a way that its time evolution leads finally the beam to the region of local regime where the self modulation becomes unstable. To study this physical circumstance, we restrict our analysis to the case of $B_0 = 0$ ($k_s = k_p$), whose Sagdeev potential corresponds to the small-dotted line in Figure \ref{F1}. It is now useful to plot it as function of the dimensionless variable $\sigma'\equiv k_p \sigma_\bot$. According to Figure \ref{F3}, the plot can be qualitatively divided into the three regions of $\sigma'$, ranging from $\sigma'= 0$ to at least $\sigma'\sim 10-10^{2}$, corresponding to strongly nonlocal (i.e., $\sigma'\ll 1$), moderately non local (i.e., $\sigma'\sim 1$) and purely local (i.e., $\sigma'\gg 1$) regimes, respectively. Therefore, the Sagdeev potential should assume different shapes in the different regions. This implies that, out of the region $\sigma' \ll 1$, the Sagdeev potential has to be determined starting with the appropriate Vlasov-Poisson-type system of equations. Since we are interested in providing a qualitative explanation of the unstable evolution of the self-modulation, we determine the analytical expression of the Sagdeev potential in the purely local region. Then, we try to approach qualitatively the moderately nonlocal region just by prolonging, one toward another, the plot of the strongly nonlocal region and the plot of the purely local region, respectively. The intersection of these two branches is illustrated in Figure \ref{F3}, where the point of intersection is denoted by $p^*$. In the $\sigma'$-space it corresponds to the value $\sigma^*$. On the scale of the potential adopted in Figure \ref{F3}, the branch of the purely local region, compared to the nonlocal one, appears practically constant, because it varies much slowly in $\sigma'$. According to the analytical treatment presented in \cite{Fedele2014a} and \cite{Fedele2014b}, these two branches are respectively given by:
\begin{equation}\label{V'1}
V_1(\sigma')=\frac{1}{2}\eta \ln\sigma'^2 +\frac{\epsilon'^2}{2\sigma'^2}\,\,,\,\,0<\sigma'\leq\sigma^*\,,
\end{equation}
\begin{equation}\label{V'1}
V_2(\sigma')=\frac{1}{2} \left(\frac{\epsilon'^2}{\sigma'^2_0}-\frac{1}{2}\frac{n_b}{n_0\gamma_0}\right)\frac{{\sigma'}_0^2}{\sigma'^2} + C\,\,,\,\,\sigma'\geq\sigma^*\,,
\end{equation}
where $\epsilon' = k_p \epsilon$, $\sigma'_0 = \sigma'(0)$, and
$$
C=\frac{1}{2}\eta \ln{{\sigma^*}^2} +\frac{1}{4}\frac{n_b}{n_0\gamma_0}\frac{{\sigma'}_0^2}{{\sigma^*}^2}
$$
is the constant which allows the continuity between the two branches. Figure \ref{F4} displays qualitatively these two brunches once a qualitative smoothing of the moderately non-local regime is done. This, of course, is only a qualitative way to connect the nonlocal region with the local one. From Figures \ref{F3} and \ref{F4} it is evident that, with initial conditions $\sigma'_0$ falling in the strongly nonlocal region (i.e., $\sigma'_0 \ll 1$) and such that the total energy of the representative point in the Sagdeev potential is $\leq C$, the variation in
$\sigma'$ are stable (periodic self-modulations). But if the initial conditions correspond to a total energy which is $ > C$, then the representative point is no longer trapped in the potential. This corresponds to a progressive increasing of $\sigma'$, i.e., to an unstable evolution of the beam envelope modulations.

\section{CONCLUSIONS AND REMARKS}
In this paper, we have described the self-modulation of a long, non laminar, relativistic electron (or positron) beam in a plasma. The process is driven by the self consistent plasma wake field excitation. The analysis has been carried out within the framework of classical description. In the strongly nonlocal regime, we have analyzed the physical conditions that shows self-modulation of the beam envelope. We have observed the self-modulation in the form of sausage-like transverse oscillations as well as the instability.
\newline
It is worthy noting that the flatness of the asymptotic region of
$\sigma'$ (i.e., purely local regime) does not depend on the sign of the difference
{\large$\left(\frac{\epsilon'^2}{\sigma'^2_0}-\frac{1}{2}\frac{n_b}{n_0\gamma_0}\right)$}. However, according to Ref. \cite{Fedele2014a}, it is easy to see that the time evolution in the purely local region is given by
\begin{equation}\label{PLR}
{\sigma'}^2(\xi')= \left(\frac{\epsilon'^2}{\sigma'^2_0}-\frac{1}{2}\frac{n_b}{n_0\gamma_0}\right)\left(\xi'-\xi^*\right)^2 +{\sigma^*}^2\,,
\end{equation}
where $\xi'= k_p\xi$ and $\xi^*$ is the timelike value of $\xi'$ such that $\sigma' (\xi^*) =\sigma^*$. Therefore, the growth of $\sigma'$ is compatible only with the positive value of  the difference {\large$\left(\frac{\epsilon'^2}{\sigma'^2_0}-\frac{1}{2}\frac{n_b}{n_0\gamma_0}\right)$}.

\end{document}